\documentclass[english]{article}
\usepackage[T1]{fontenc}
\usepackage[latin9]{inputenc}

\makeatletter
\newcommand{\lyxaddress}[1]{
\par {\raggedright #1
\vspace{1.4em}
\noindent\par}
}

\makeatother

\usepackage{babel}
\begin{document}

\title{A new approach to Gravity}

\author{S. B. Faruque}

\maketitle

\lyxaddress{Department of Physics, Shahjalal University of Science and Technology,
Sylhet, Sylhet 3114, Bangladesh}
\begin{abstract}
Beginning with a decomposition of the Newtonian field of gravity,
I show that four classical color fields can be associated with the
gravitational field. The meaning of color here is that these fields
do not add up to yield the Newtonian gravitational field, but the
forces and potential energies associated with them add up to yield
the Newtonian force and potential energy, respectively.These four
color fields can have associated magnetic fields as in linearzied
gravity. Thus we envisage a theory where four sets of Maxwellian equations
would prevail. A quantum gravity theory with four spin 1 fields can
thus be envisaged.
\end{abstract}

\section{Introduction}

Gravitational force is attractive despite there is nothing opposite
in the masses of the attracting bodies. Newtonian gravity is the low
energy version of the general theory of relativity and linearized
gravity stands in between them. In linearized gravity, attracting
bodies, the source body and the test mass, are assumed to have opposite
masses. In this theory, a gravitomagnetic charge is attributed to
both the source and the test mass. Both gravitoelectric and gravitomagnetic
charges of the test and source bodies are assumed to have opposite
sign. This is seemingly justified, but conceptually unacceptable,
because, there is so far no evidence of difference in what we call
mass is found. Therefore, if the source is of positive gravitoelectric
charge, then the test body should also have the same positivity. If
the gravitomagnetic charge is negative in one, then it should be the
same in the other. Beginning with this conceptual departure from linearized
gravity, I formulate the Newtonian portion of gravity in a way that
can be extended to a quantum theory. Here, I decompose the Newtonian
field of gravity into four parts with which gravitoelectric and gravitomagnetic
charges interact. The fields are not additive, but the forces and
potential energies arising from the four fold interaction add up to
yield the same force and potential energy that we find in Newtonian
gravity. I present the fields and calculation of the forces and potential
energies in this new approach to gravity in the next section.

\section{Fields and forces of gravity}

For slowly moving gravitational sources, the field equations of gravity
resemble those of electrodynamics{[}1{]}. Hence, a formal theory called
gravitoelectromagnetism has evolved in course of time. In this theory,
the standard formulae of electrodynamics are applicable, except that
for a source of inertial mass M, the gravitoelectric charge is M and
the gravitomagnetic charge is 2M. On the other hand, for a test particle
of mass m, the gravitoelectric charge is -m and the gravitomagnetic
charge is -2m. The source and the test particle have opposite charges
to ensure that gravity is attractive {[}2-3{]}. But, as we know very
well that there are no evidences in favor of difference between source
and test bodies regarding their that property which is called mass,
one feels uncomfortable with such a difference in signs of the charges
of source and test bodies. To avoid the seemingly justified, but conceptually
unbearable unparallel convention of source and test particle charges,
we have come across a new way of making gravity attractive always
without assigning charges to the sources and test partticles that
do not match.

In our formulation, we shall treat the source body of mass M as possessing
gravitoelectric charge $q_{E}=M$ and gravitomagnetic charge $q_{B}=-2M$.
The same convention will be used for the test particle; so the test
particle possesses gravitoelectric charge $q_{E}=m$ and gravitomagnetic
charge $q_{B}=-2m$. The source body will be assumed to produce four
fields that are associated with the gravitoelectric and gravitomagnetic
charges. Both of these charges produce fields that act on the test
particle charges depending on the nature of the charges. That means,
if we call $\vec{E}_{e}$ the electric field due to gravitoelectric
charge, then $\vec{E}_{b}$ is the electric field due to gravitomagnetic
charge. $\vec{E_{e}}$ has to colors, one, $\vec{E}_{ee}$ acts only
on electric charge, and the other, $\vec{E}_{eb}$ acts only on magnetic
charge. In the same way, $\vec{E}_{be}$ acts only on electric charge
and $\vec{E}_{bb}$ acts only on magnetic charge. Hence, we find four
fields. We assume the expressions for these color fields, in some
analogous way to ordinary gravity, as (G=1):
\begin{equation}
\vec{E}_{ee}=\frac{M}{r^{2}}\hat{r},
\end{equation}
\begin{equation}
\vec{E}_{bb}=-\frac{2M}{r^{2}}\hat{r},
\end{equation}
\begin{equation}
\vec{E}_{eb}=\frac{3}{2}\frac{M}{r^{2}}\hat{r},
\end{equation}
\begin{equation}
\vec{E}_{be}=-\frac{3M}{r^{2}}\hat{r},
\end{equation}

where the source body of mass M is assumed to be located at the origin
of coordinate system and r is the radial coordinate.The characteristics
of these fields, as evident from the expressions, are that gravitoelectric
charge produce diverging fields and gravitomagnetic charge produce
converging fields. Also, the cross fields are 3/2 times stronger than
the direct fields.

The test particle of gravitoelectric charge m and gravitomagnetic
charge (-2m) experience four forces given by the symmetric prescription,
\begin{equation}
\vec{F}=q\vec{E}
\end{equation}
 as given by the following expressions:
\begin{equation}
\vec{F}_{1}=(m)\vec{E}_{ee}=\frac{Mm}{r^{2}}\hat{r,}
\end{equation}
\begin{equation}
\vec{F}_{2}=(-2m)\vec{E}_{bb}=\frac{4Mm}{r^{2}}\hat{r,}
\end{equation}
\begin{equation}
\vec{F}_{3}=(-2m)\vec{E}_{eb}=-\frac{3Mm}{r^{2}}\hat{r,}
\end{equation}
\begin{equation}
\vec{F}_{4}=(m)\vec{E}_{be}=-\frac{3Mm}{r^{2}}\hat{r.}
\end{equation}

We note that
\begin{equation}
\vec{E}_{resultant}=\vec{E}_{ee}+\vec{E}_{bb}+\vec{E}_{eb}+\vec{E}_{be}=-\frac{5}{2}\frac{M}{r^{2}}\hat{r}\neq\vec{E}_{Newtonian}.
\end{equation}

However,
\begin{equation}
\vec{F}_{resultant}=\vec{F}_{1}+\vec{F}_{2}+\vec{F}_{3}+\vec{F}_{4}=-\frac{Mm}{r^{2}}\hat{r}=\vec{F}_{Newtonian}.
\end{equation}

Hence, the fields $\vec{E}_{ee}$, $\vec{E}_{bb}$ etc. are colored;
they do not add up to the Newtonian field, but the forces due to the
four color fields on the gravitoelectric and gravitomagnetic charges
of the test particle add up to the net force as equal to what we know
as the Newtonian gravitational force.

The resulting potential energies of the test particle is given by
the prescription $\vec{F}=-\vec{\nabla U}$, U being the potential
energy, as
\begin{equation}
U_{1}=\frac{Mm}{r},
\end{equation}
\begin{equation}
U_{2}=\frac{4Mm}{r},
\end{equation}
\begin{equation}
U_{3}=-\frac{3Mm}{r},
\end{equation}
\begin{equation}
U_{4}=-\frac{3Mm}{r}.
\end{equation}

and their sum
\begin{equation}
U_{resultant}=-\frac{Mm}{r}=U_{Newtonian}.
\end{equation}

Hence, like the force, the potential energy is colorless and add up
to yield the correct resultant as given by the Newtonian potential
energy. The four potentials in which the test particle finds itself
are given by $\Phi=\frac{U}{q},$q being either $q_{E}$ or $q_{B}$,
as
\begin{equation}
\Phi_{ee}=\frac{M}{r},
\end{equation}
\begin{equation}
\Phi_{bb}=-\frac{2M}{r},
\end{equation}
\begin{equation}
\Phi_{eb}=\frac{3}{2}\frac{M}{r},
\end{equation}
\begin{equation}
\Phi_{be}=-\frac{3M}{r}.
\end{equation}

We note that 
\begin{equation}
\Phi_{resultant}=-\frac{5}{2}\frac{M}{r}\neq\Phi_{Newtonian}.
\end{equation}

Therefore, the effective force due to all the fields, experienced
through the gravitoelectric and gravitomagnetic charges, on the test
particle is exactly the same as the Newtonian force on the particle.
The fields themselves do not superpose to yield a resultant equivalent
to the Newtonian field. Hence, the fields are color-sensitive. At
the end, we regain the same force and potential energy experienced
by the test particle as if it has nothing extra of the Newtonian formulation.
After all, Newtonian force and potential energy are the reality at
the low energy scale. In this way, we have gained a compromise between
linearized gravity's two types of charges and Newtonian gravity's
attractive force, without giving up the parallel between source and
test particle masses (charges).

\section{Discussion}

We have shown that the Newtonian gravitational force between two bodies
can be obtained using linearized gravity's two types of masses, namely,
gravitoelectric and gravitomagnetic charges. In linearized gravity,
conventionally the source body's charges and test particle charges
are opposite in sign. But, here we ascribe to the source body and
test paticle charges of the same sign. We then associate four color
fields with the source mass and compute the individual forces experienced
by the test particle due to its gravitoelectric and gravitomagnetic
charges. The individual forces add up to exactly the Newtonian gravitational
force. The potential energies also add up to the Newtonian potential
energy. But, the fields and potentials do not superpose, i.e., they
are colorful. Thus, in this new approach to gravity, four color fields
are responsible for the gravitational force. Hence, we can envisage
that four sets of Maxwellian equations would prevail when the theory
is extended using four magnetic fields ascribed with the four electric
color fields. Finally, a quantum gravity theory with four spin 1 fields
can be envisaged which would complete a theory of gravity.

\end{document}